\documentclass[]{aastex631}

\shorttitle{Spatially dependent correction of Gaia EDR3 parallax zero-point offset}
\shortauthors{Wang C. et al.}
\graphicspath{{./}{figures/}}

\begin{document}

\title{A spatially dependent correction of Gaia EDR3 parallax zero-point offset based on 0.3 million LAMOST DR8 giant stars}

\author{Chun Wang}
\altaffiliation{Corresponding author (wchun@tjnu.edu.cn)}
\affiliation{Tianjin Astrophysics Center, Tianjin Normal University, Tianjin 300387, People's Republic of China.}
\author{Haibo Yuan}
\affiliation{Department of Astronomy, Beijing Normal University, Beijing 100875, People’s Republic of China.}

\author{Yang Huang}
\affiliation{South-Western Institute for Astronomy Research, Yunnan University, Kunming, Yunnan 650091, People’s Republic of China.}

\begin{abstract}

We have studied the zero-point offset of Gaia early Data Release 3 (EDR3) parallaxes based on a sample of   0.3 million giant stars built from the LAMOST data with distance accuracy better than  8.5\%.  
The official parallax zero-point corrections largely reduce the global offset in the Gaia EDR3 parallaxes: the global parallax offsets are $-$27.9 $\mu$as and $-$26.5 $\mu$as (before correction) and $+$2.6 $\mu$as and $+$2.9 $\mu$as (after correction) for the five- and six-parameter solutions, respectively.
 The  bias of the raw parallax measurements is significantly dependent on the $G$ magnitudes, spectral colors, and positions of stars. The official parallax zero-point corrections could reduce parallax bias patterns with $G$ magnitudes, while could not fully account the patterns in the spaces of the spectral colors and positions.  In the current paper, a spatially dependent parallax zero-point correction model for Gaia EDR3 five-parameter solution in the LAMOST footprint is firstly provided with the advantage of huge number of stars in our sample. 
\end{abstract}



\section{Introduction} \label{sec:intro}

Gaia Early Data Release\,3 \citep{Gaia_Collaboration2021}
  have released  astrometric and photometric data for over 1.8 billion sources based on observations from the first 34 months (from 2014 July to 2017 May) by the European Space Agency's Gaia mission \citep{Gaia_Collaboration2016}.  Among of them,  1.468 billion sources  have full astrometric data, including position, parallax, and proper motions. The data set of Gaia EDR3 is widely used in the fields of the stellar and Galactic astrophysics.   
 
 In Gaia EDR3,  the  typical uncertainties of parallaxes  are 0.03--1.4 mas for stars with $15 < G < 21$\,mag \citep{Gaia_Collaboration2021}.  The systematic parallax errors are inevitable because of the imperfections in the instruments and data processing \citep{Lindegren2021}, which will produce large systematic distance errors especially for more distant stars. 
 Thus, investigating the systematic bias of the Gaia EDR3 parallax is important for its further applications. 
\cite{Lindegren2021b} have constructed a parallax zero-point correction model, which is a function of  the $G$-band magnitude, spectral shape (colors), and ecliptic latitude of the sources. This correction model is  based on  quasars distributed across the entire sky, binary stars, and stars in the Large Magellanic Cloud (LMC).  The correction models are different for sources with five- and six-parameter solutions. However, the correction models are mostly  appropriate for only sources having similar magnitudes and colors to quasars.  Thus the independent test of parallax zero-point offsets of Gaia EDR3 using Galactic stars is needed.  Several papers have also contributed to the parallax zero-point offset of Gaia EDR3 estimates using different tracers, including quasars, RR Lyrae stars, red clump stars, red giant stars and binaries \citep{Bhardwaj2021,Liao2021,Groenewegen2021,ElBadry2021,Stassun2021,Zinn2021,Huang2021,Ren2021}. 
Amongst of them, \cite{Huang2021} and \cite{Ren2021}  estimate the independent  parallax zero-point offsets of Gaia EDR3 using $\sim$ 0.07 million LAMOST primary red clump (PRC) stars and $\sim$ 0.11 million W Ursae Majoris (EW)-type eclipsing binary systems  in a wide  magnitude and color ranges with the advantage of large stellar sample and their accurate measurements of distance.  They find  $\sim -27 \, \mu \rm as$ and $\sim +4 \,\mu \rm as$ global zero-point offset of Gaia EDR3 parallaxs before and after correction, respectively.  The official parallax zero-point model of \cite{Lindegren2021b} could reduce the global bias in the Gaia EDR3 parallax. After correction, the parallax bias variations with spectral colors and positions still exist.   Thus, it is useful  to investigate the  zero-point offset in Gaia EDR3 parallaxes, especially the parallax zero-point offset variations with the positions.

Here, we use the sources targeted by LAMOST  to  investigate the zero-point offset in Gaia EDR3 parallaxes.  We will introduce our data set in Section\,2. The main results are presented and discussed
in Section 3.   Finally, Section 4 presents a summary of our results.

\section{Data}
We use the data of the value-added catalog for LAMOST DR8 low resolution spectra (Wang et al. 2022, accepted) to test the zero-point offset of Gaia EDR3.  The catalog contains  accurate photometric distances for about 7.1 million spectra of 5.17 million unique stars with spectral signal-to-noise ratios (SNRs) higher than 10 obtained by the Large Sky Area Multi-Object Fibre Spectroscopic Telescope (LAMOST) Galactic spectroscopic surveys \citep{deng-legue, Zhao2012, liu-lss-gac}.   The photometric distance  of individual stars observed by LAMOST are estimated with distance modulus method using the 2MASS \citep{Skrutskie2006} $K_{s}$ band apparent magnitudes, interstellar extinctions and 2MASS $K_{s}$  band absolute magnitudes ($M_{Ks}$) derived directly from the LAMOST spectra using neural network models.  The photometric distance is accurate to 8.5\% for  stars with spectral SNRs larger than 50.  Especially, the photometric distance show small dependence on the apparent magnitudes, positions of stars, absolute magnitudes and spectral colors.  It is a suitable dataset to test  the zero-point offset in Gaia EDR3 parallaxes. 

\subsection{Sample selection}
We cross-match the value-added catalog with Gaia EDR3 and 2MASS, and use the  following criteria to select  stellar sample to test the zero-point offset of Gaia EDR3 parallax:
\begin{itemize}
\item LAMOST spectral signal to noise ratio (SNR) $\geq$50, parallaxes  smaller than 1.5\,mas, surface gravity $\log g \leq 3.8$, and the uncertainty of 2MASS $K_{s}$ band apparent magnitude smaller than 0.03\,mag.
\item The renormalized unit weight error (RUWE) $\leq$ 1.4. 
\item An effective wavenumber 1.1$\leq \nu_{\rm eff}\leq$1.9   for the five-parameter solution  and 1.24 $\leq$ pseudocolor $\leq 1.72$ for six-parameter solution. 
\item The spectral color $0.3 < B_{p}-R_{p} < 1.65$ mag. 
\end{itemize}

\begin{figure*}
\centering
\includegraphics[width=5.5in]{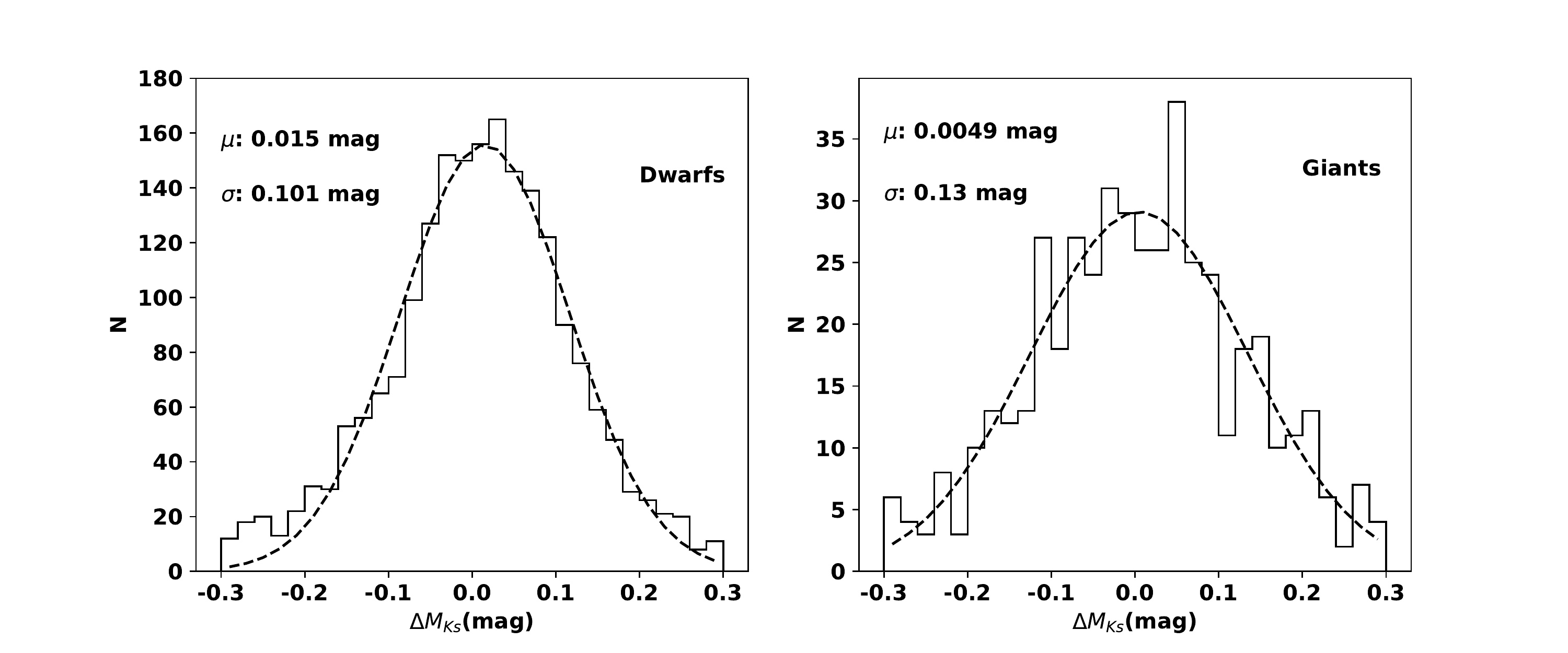}
\caption{The distributions of  $M_{Ks}$ difference  between  $M_{Ks}$ coming from Wang et al. (2022, accepted) ($M_{Ks}^{W}$)  and that estimated using the distance of \cite{Baileredr3} ($M_{Ks}^{Gaia}$) of the test sample in Wang et al. (2022, accepted). The left and right panels are the results of dwarfs and giants, respectively.  The black dashed line represent Gaussian fits for the difference distributions.  The mean values and standard deviations of the two distributions are labeled in the top left corner of each panel. The $\Delta M_{Ks}$ is the  $M_{Ks}^{Gaia}-M_{Ks}^{W}$.}   
\label{train_dwarf_giant}
\end{figure*}

Only  distant giant stars are selected for this study as shown in the first selection criterion. The systematic difference  between $M_{Ks}$ coming from Wang et al. (2022, accepted)
and that estimated using the distance of \cite{Baileredr3} of our test sample are 0.015 mag and 0.0049 mag for dwarf and giant stars as shown in Fig.\ref{train_dwarf_giant}, respectively (Wang et al. 2022, accepted).  
Because the typical parallaxes of the dwarf and giant stars in our sample are 1.5 and 0.45 mas, respectively, these systematic differences in $M_{Ks}$ will typically produce a systematic parallax difference of about $\sim 10.3 \, \mu \rm as$ for the dwarf stars, and  $\sim 1.0 \,\mu \rm as$ for the giants. Furthermore,  the effect of binary stars  on the estimated absolute magnitudes of dwarf stars is larger than that of giant stars. 
Thus,  we only select distant giant stars as tracers to check the zero-point offset of Gaia EDR3 parallax in the current work. 
 The RUWE is a  quality indicator of Gaia EDR3 data, given by the renormalized square root of the reduced chisquare of the Gaia astrometric solution.  Larger values indicate that the astrometric solution does not completely describe the source motion \citep{Lindegren2021,Fabricius2021}. The stars with large value of RUWE are also removed.   The  effective wavenumber and pseudocolor represent the color information of the observed targets. Besides, the stars with  $B_{p}-R_{p} \leq 0.3$\,mag or $B_{p}-R_{p} \geq1.65$\,mag  are also removed in order to make sure that our training sample could cover all kinds of stars in our giant sample (one can see Section\,2.2 for more details).  After the above cuts, we select about 268,003 and 5,070 LAMOST giant stars with five-parameter  and six-parameter solutions in Gaia EDR3, respectively.

\subsection{The  parallax quality of our giant sample}

\begin{figure*}
\centering
\includegraphics[width=5.5in]{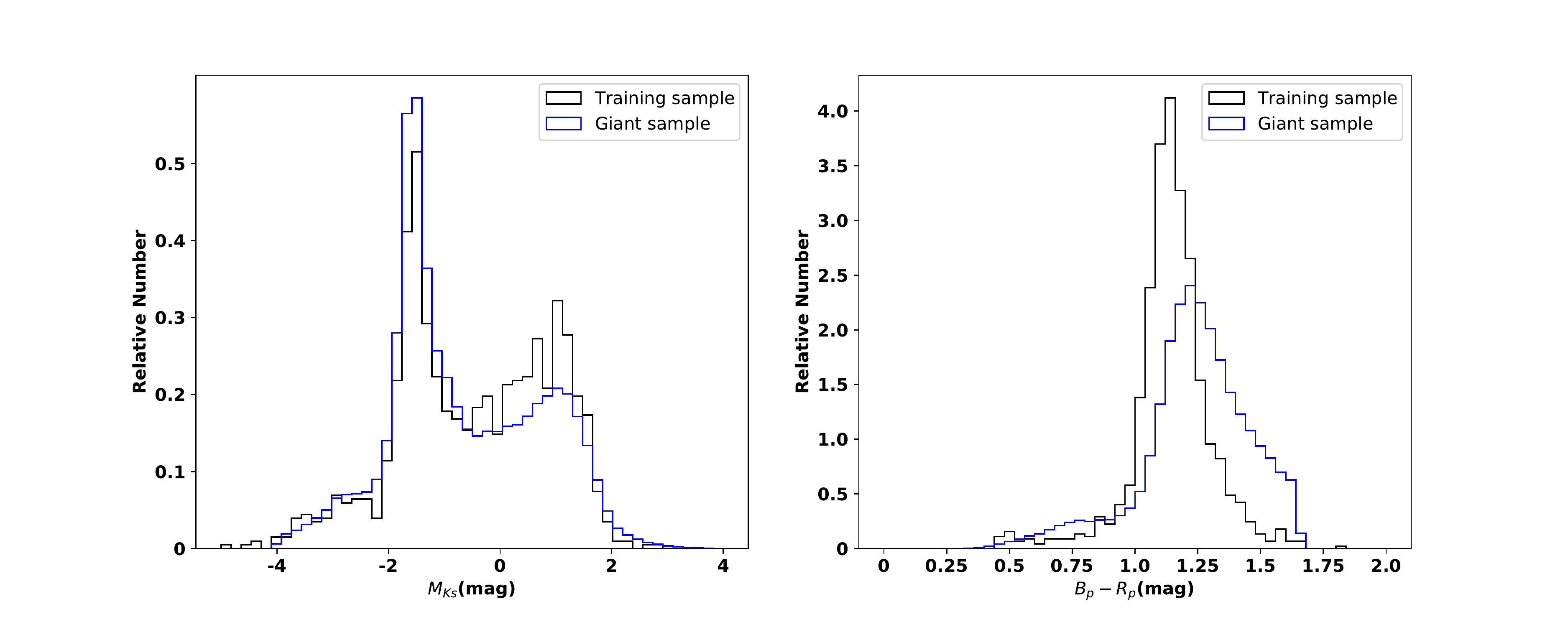}
\caption{The absolute magnitude (left panel) and spectral color (right panel)  distributions of the  LGMWAS  training sample (black lines) and our final giant sample (blue lines). }
\label{compare_giant_bprp}
\end{figure*}

 As shown in Wang et al. (2022, accepted), the photometric distance are estimated with  the distance modulus method using the 2MASS  $K_{s}$ band apparent magnitudes, interstellar extinctions and $M_{Ks}$ derived directly from the LAMOST spectra. Amongst of them, the $M_{Ks}$ are derived using neural network models, which are built up using the LGMWAS (LAMOST--Gaia EDR3--2MASS--WISE--APASS--SDSS ) common stars as training set.  In the LGMWAS training sample, the absolute magnitudes are estimated using the geometrical Gaia EDR3 distance \citep{Baileredr3}, 
based on the Gaia EDR3 parallaxes corrected according to \cite{Lindegren2021b}. We can nevertheless use the distances in Wang et al. (2022, accepted) to test the parallax bias in Gaia EDR3 because the mean parallax of the LGMWAS training set ($\sim$ 2 mas) is much larger than the mean parallax in our giant sample ($\sim$ 0.45 mas). Our sample is therefore 4--5 times more sensitive to a parallax bias than the training set, which breaks the degeneracy, considering also that the two samples cover the same ranges of absolute magnitude and spectral color as shown in Fig.\,\ref{compare_giant_bprp}.
  
In Wang et al. (2022, accepted), we only select stars with small extinctions ($E (B-V) < 0.02$\,mag) as training stars. Almost all of  our giant stars  have larger extinction (with a median value of $E (B-V) \sim 0.09$\,mag, estimated using ``star-pair" method of \cite{ Yuan2013}) and extinction errors (with a typical value of 0.02\,mag) than that of training stars,  which will produce larger errors of the estimated photometric distance. 
The uncertainties of 0.006\,mag for the 2MASS $K_{s}$ band apparent magnitude  and 1.2 $\mu \rm as$ for the the estimated parallax  will be caused by the typical extinction error.  This  parallax error is very small even if it is a systematic error. 
Thus, the variations of estimated parallax zero-point offset  with spectral color, apparent magnitude and spatial positions will be less affected considering also the huge number of stars in our giant sample. 
In conclusion, although the estimates of parallax in our giant sample are based on the  geometrical Gaia EDR3 distance of \cite{Baileredr3}, we could use the sample to test the parallax bias of Gaia EDR3 and its dependence on the spectral color, apparent magnitude and spatial positions.  

\begin{figure*}
\centering
\includegraphics[width=3.5in]{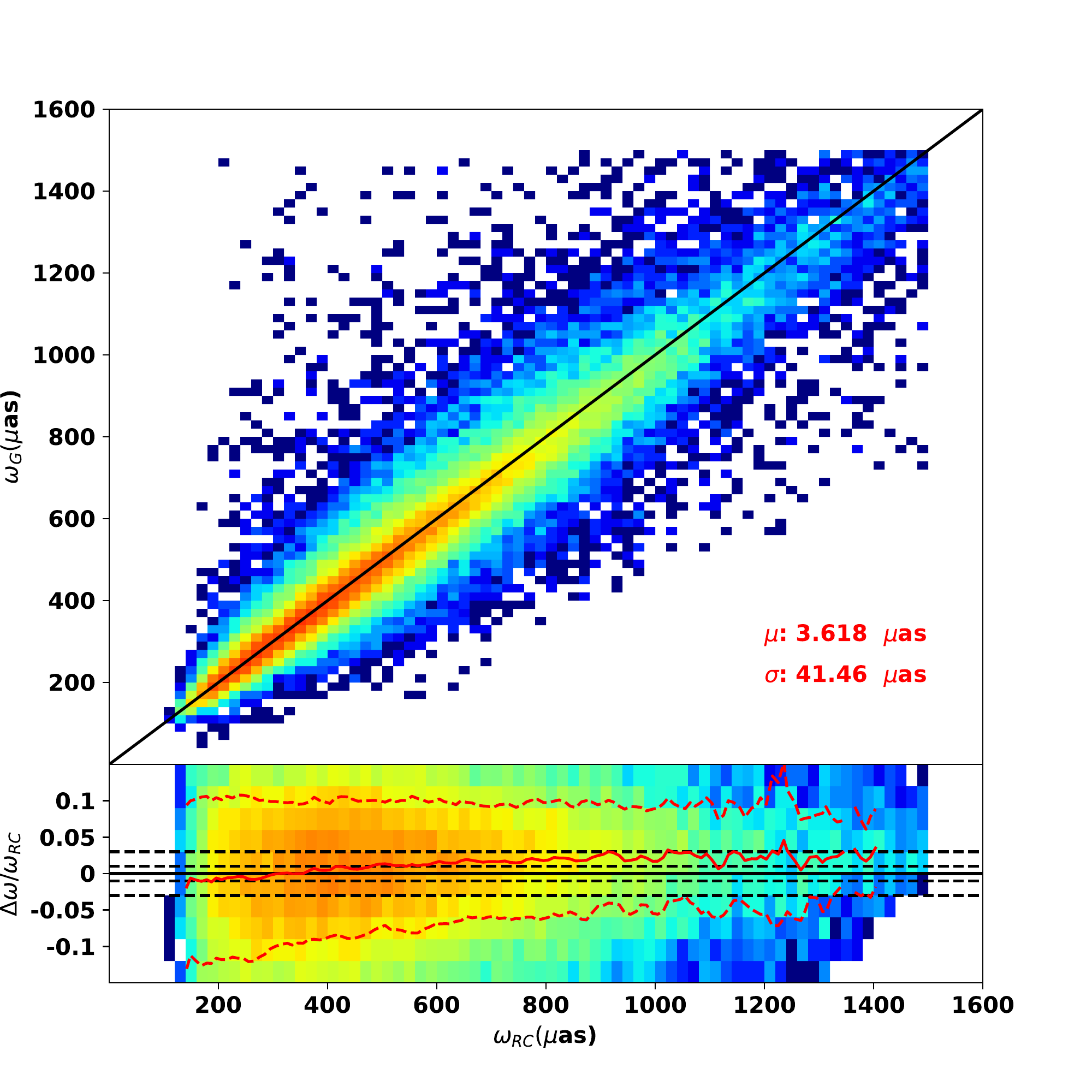}
\caption{Top panel: the comparison between  our parallax  and PRC parallax.  Bottom panel: the variations of  relative parallax difference  between our parallax  and PRC parallax  with the PRC parallax. The  black dashed lines in the bottom panel are zero minus/plus 1\% or 3\%.}
\label{compare_parallax_rc}
\end{figure*}

PRC stars are standard candles, their absolute magnitudes and distances could be accurately estimated \citep{Cannon1970, Paczynski1998, Bovy2014,Huang2015, Wan2015, Chen2017,Huang2020}.  \cite{Huang2020} have selected $\sim 140,000$ PRCs from LAMOST, the distances of which   could be accurate to 5\% for stars with spectral SNRs larger than 50.  We cross-match our giant sample and PRC sample of \cite{Huang2020}, and select 49,459 common stars. 
Fig.\,\ref{compare_parallax_rc} shows the comparison of our parallax ($\omega_{G}$) and PRC parallax ($\omega_{RC}$) and the parallax difference variations with the PRC parallax. The global systematic difference of the two kinds of parallax is  3.6\,$\mu\rm as$, which is slightly larger than the claimed typical parallax systematic difference ($\sim 1.0 \mu\rm as$) between our parallax   and official zero-point offset corrected Gaia EDR3 parallax as discussed in the Section 2.1.  It is reasonable considering the systematic difference between  the corrected Gaia EDR3 parallax and the parallax of PRC sample. The standard  deviation of the parallax difference is $\sim 41 \, \mu as$, which is also good enough.  
From Fig.\,\ref{compare_parallax_rc}, we can find a  trend of parallax difference with the parallax.  
The parallax difference are $-1$\%, 1\% and 3\% at $\omega_{RC} \sim 200\,\mu\rm as$ (corresponding parallax error is\,$\sim -2\,\mu \rm as$), $\omega_{RC} \sim 450\,\mu\rm as$ (corresponding parallax error is $\sim +4.5\,\mu \rm as$) and $\omega_{RC} \geq 1000\,\mu\rm as$ (corresponding parallax error is $\geq +30\,\mu\rm as$), respectively.  Considering the typical parallax of  our giant stars is  $\sim 450\,\mu\rm as$, the several $\mu\rm as$ systematic error of our giant sample and its variations with parallax will produce small systematic error of our final estimated parallax zero-point offset, which are supported by the similar global parallax zero-point offsets of us and other previous works discussed in Section\,3.  Besides, the stars with different parallax have similar spatial distributions, which suggest that the relative spatial variations of the estimated parallax zero-point offset  will not be affected by the trend.  
In conclusion, our giant sample is a good sample to  test the Gaia EDR3 parallax zero-point offset, especially its variations with apparent magnitude, spectral color and spatial positions. 
 
 In the current paper,  $\omega_{\rm EDR3}$, $\omega_{\rm EDR3}^{\rm corr}$ and $\omega_{\rm G}$ (here 'G' mean giant stellar sample) are the Gaia EDR3 parallax, Gaia EDR3 parallax with zero-point correction using the model of \cite{Lindegren2021b} and the parallax given by photometric  distance of our LAMOST giant stars, respectively. 

\section{Results and Discussions}

\begin{figure*}
\centering
\includegraphics[width=5.5in]{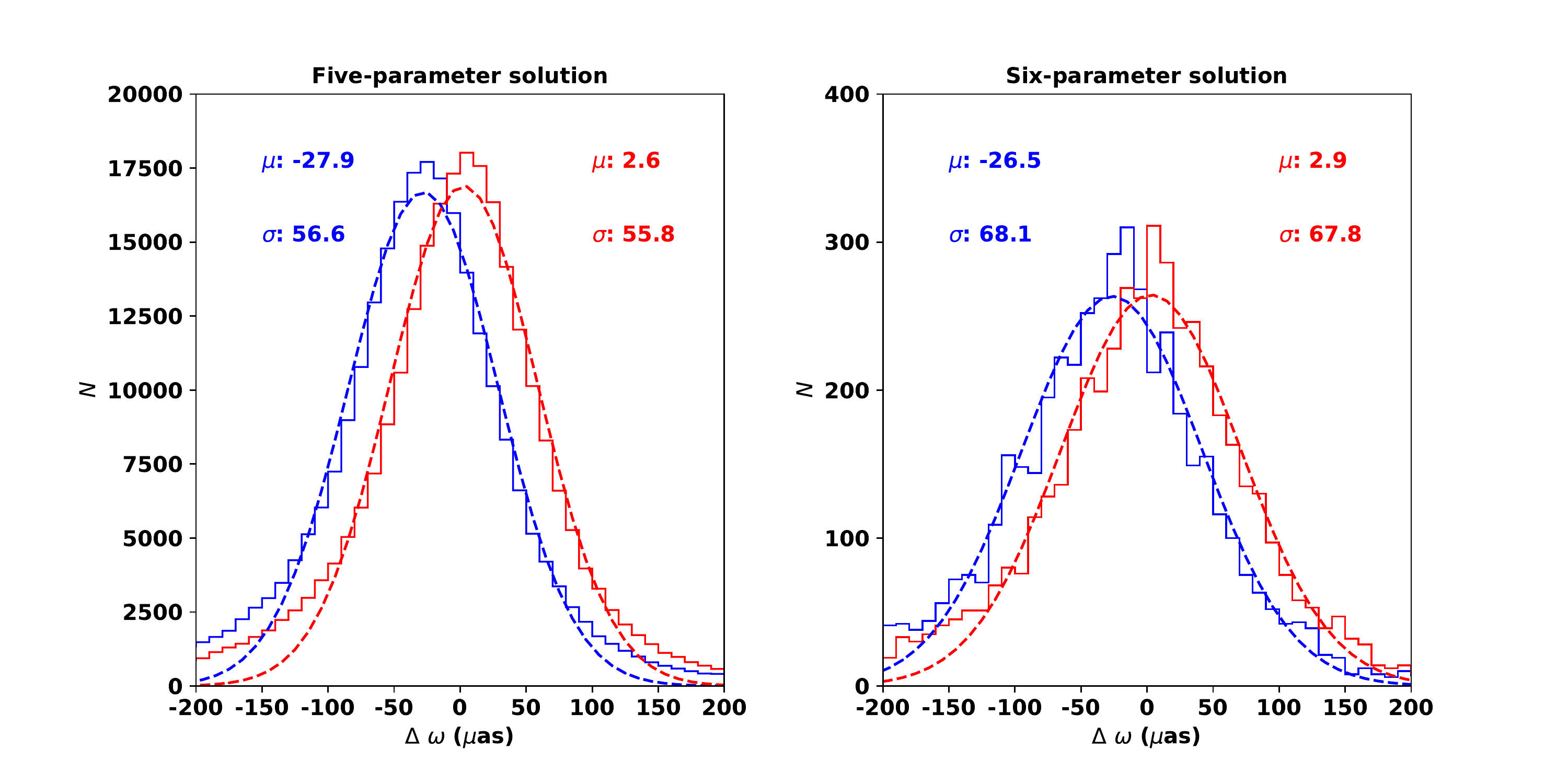}
\caption{Differences distributions of parallaxes between the LAMOST giant stellar sample  and  Gaia EDR3 five-parameter solution (left panel) or six-parameter solution (right panel). The blue and red histograms represent the $\omega_{\rm EDR3}-\omega_{\rm G}$ and $\omega_{\rm EDR3}^{\rm corr}-\omega_{\rm G}$ distributions, respectively.  The blue and red dashed lines represent Gaussian fits for the two parallax difference distributions, respectively.  The mean values and standard deviations of the two distributions are labeled in the top corner using their corresponding colors.}
\label{gaiaedr3_zeropoint_all}
\end{figure*}

Using the selected giant stars, we compare the  parallaxes derived from our photometric distances ($\omega_{\rm G}$) with those from Gaia EDR3 ($\omega_{\rm EDR3}$)  for  the five- and six-parameter solutions. As shown in  Fig.\,\ref{gaiaedr3_zeropoint_all},   the median offsets of Gaia EDR3 parallaxes  are respectively $-27.9$ $\mu \rm as$ and  $-26.5$ $\mu \rm as$ for the five- and six-parameter solutions, which are slightly larger than the value of $-17 \mu \rm as$ derived  from distant quasars \citep{Lindegren2021b}.  Our estimated median offsets are consistent with that of \citet[$\sim -26 \, \mu \rm as$ for both five- and six-parameter solutions]{Huang2021}  and \citet[$\sim -28$ and $-25 \, \mu \rm as$ for respectively five- and six-parameter solutions]{Ren2021}.   The median values of the
parallax difference between $\omega_{\rm EDR3}^{\rm corr}$ and $\omega_{\rm G}$  are only $+2.6$ and
$+2.9$ $\mu \rm as$ for the five- and six-parameter solutions, respectively. The negligible positive values of median offsets suggest that the official parallax zero-point correction model \citep{Lindegren2021b} significantly  reduces the  global bias of Gaia EDR3 parallaxes.
 
\begin{figure*}
\centering
\includegraphics[width=5.5in]{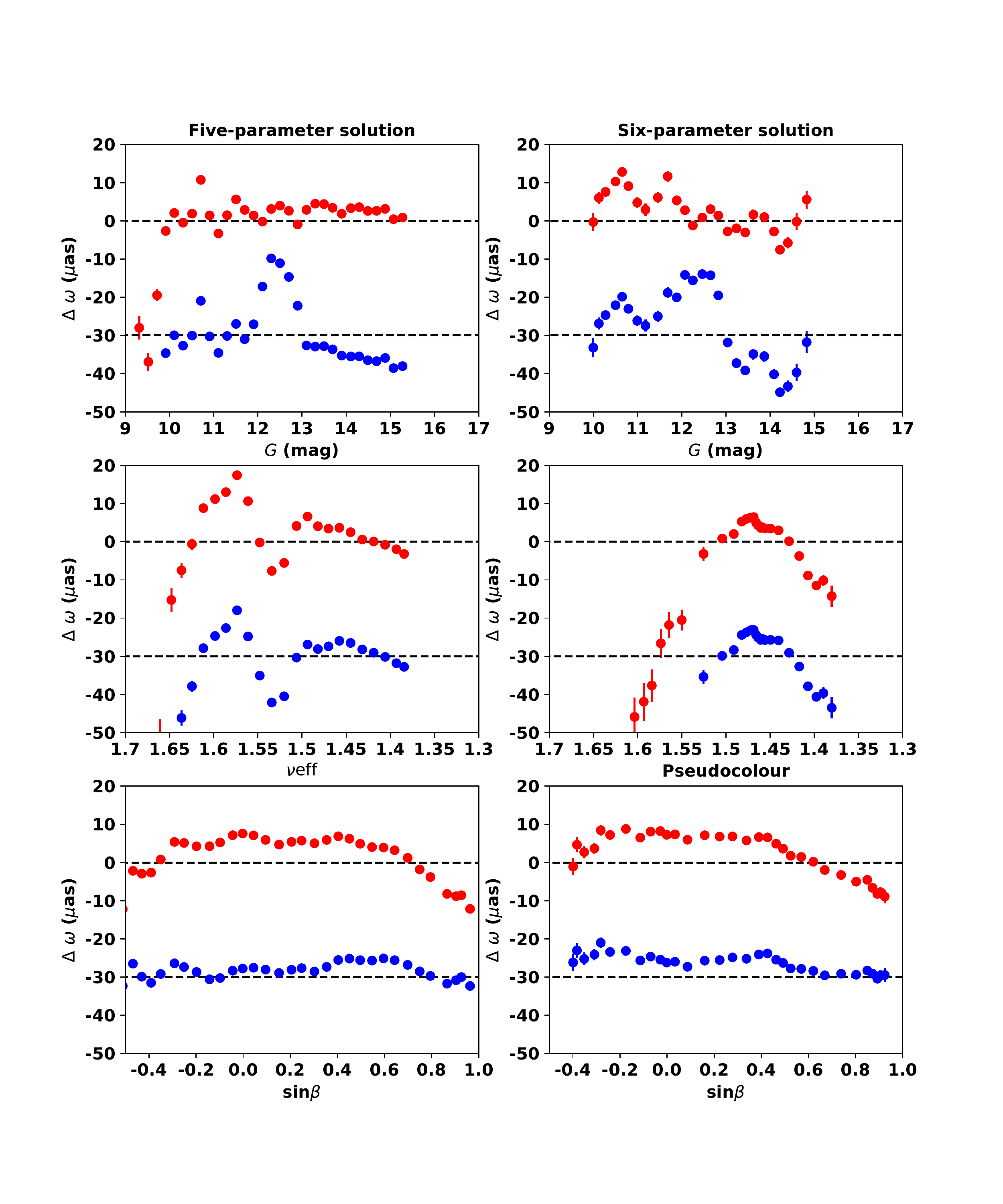}
\caption{Parallax differences  between our LAMOST  giant sample and  Gaia EDR3 five-parameter solution (left panel) or Gaia EDR3 six-parameter solution (right panel),  as a function of $G$ band magnitudes (top panels),  color information (middle panels: $\nu_{\rm eff}$ for five-paramter solution, psceudocolour for six-paramter solution) and ecliptic latitude sin$\beta$ (bottom panels).  The blue and red points represent the ($\omega_{\rm EDR3}-\omega_{\rm G}$) and ($\omega_{\rm EDR3}^{\rm corr}-\omega_{\rm G}$), respectively.  In each bin, the number of stars  is no less than 100. The black dashed
lines in each panel mark the differences of  $-$30 $\mu \rm as$ and 0 $\mu \rm as$.}
\label{zeropoint_para_gveffsinb}
\end{figure*}

 As discussed in \cite{Lindegren2021b}, \cite{Huang2021} and \cite{Ren2021}, the bias of Gaia EDR3 parallaxes is dependent on apparent magnitude ($G$ band), effective wavenumber $\nu_{\rm eff}$/pseudocolor and ecliptic latitude (sin$\beta$).  With the advantage of large number of stars, large and continue coverage in the Galactic plane and wide range of colors and magnitudes of our giant sample, we test the main dependencies  of the Gaia EDR3 parallaxes bias. 
Fig.\,\ref{zeropoint_para_gveffsinb} shows the  parallax bias of $\omega_{\rm EDR3}-\omega_{\rm G}$ and $\omega_{\rm EDR3}^{\rm corr}-\omega_{\rm G}$, as a function of G magnitude, effective wavenumber $\nu_{\rm eff}$/pseudocolor,
ecliptic latitude sin$\beta$ for the five- and six-parameter solutions.   From Fig.\,\ref{zeropoint_para_gveffsinb}, we find that the $\omega_{\rm EDR3}-\omega_{\rm G}$ are almost $ -30\,\mu \rm as$, the $\omega_{\rm EDR3}^{\rm corr}-\omega_{\rm G}$ are almost $ 0\,\mu \rm as$, which is similar with the result shown in Fig.\,\ref{gaiaedr3_zeropoint_all}. 

From Fig.\,\ref{zeropoint_para_gveffsinb}, we can find significant patterns  of
$\omega_{\rm EDR3}-\omega_{\rm G}$ as a function of $G$ magnitude for both five- and six-parameter solutions. 
The ``hump-like" feature in  $10 < G < 11$ and $12 < G < 13$ are found for  both five- and six-parameter solutions, which is similar with the results of \cite{Lindegren2021b} and \cite{Huang2021}.    The $\omega_{\rm EDR3}^{\rm corr}-\omega_{\rm G}$  are almost equal to zero and show smaller variations with $G$ magnitude.   The results suggest that the official parallax zero-point correction could reduce parallax bias patterns with G magnitudes. 
 The $\omega_{\rm EDR3}-\omega_{\rm G}$ and $\omega_{\rm EDR3}^{\rm corr}-\omega_{\rm G}$ as a function of $\nu_{\rm eff}$/ $\rm pseudocolour$  show several patterns for five- or six-parameter solutions, respectively. 
 The results suggest that the official parallax zero-point correction could not reduce parallax bias patterns with spectral colors.  
 For the $\omega_{\rm EDR3}-\omega_{\rm G}$  and $\omega_{\rm EDR3}^{\rm corr}-\omega_{\rm G}$ 
  variations with sin$\beta$,  we do not find significant trend  for both five- and six-parameter solutions.  
 
\begin{figure*}
\centering
\includegraphics[width=5.5in]{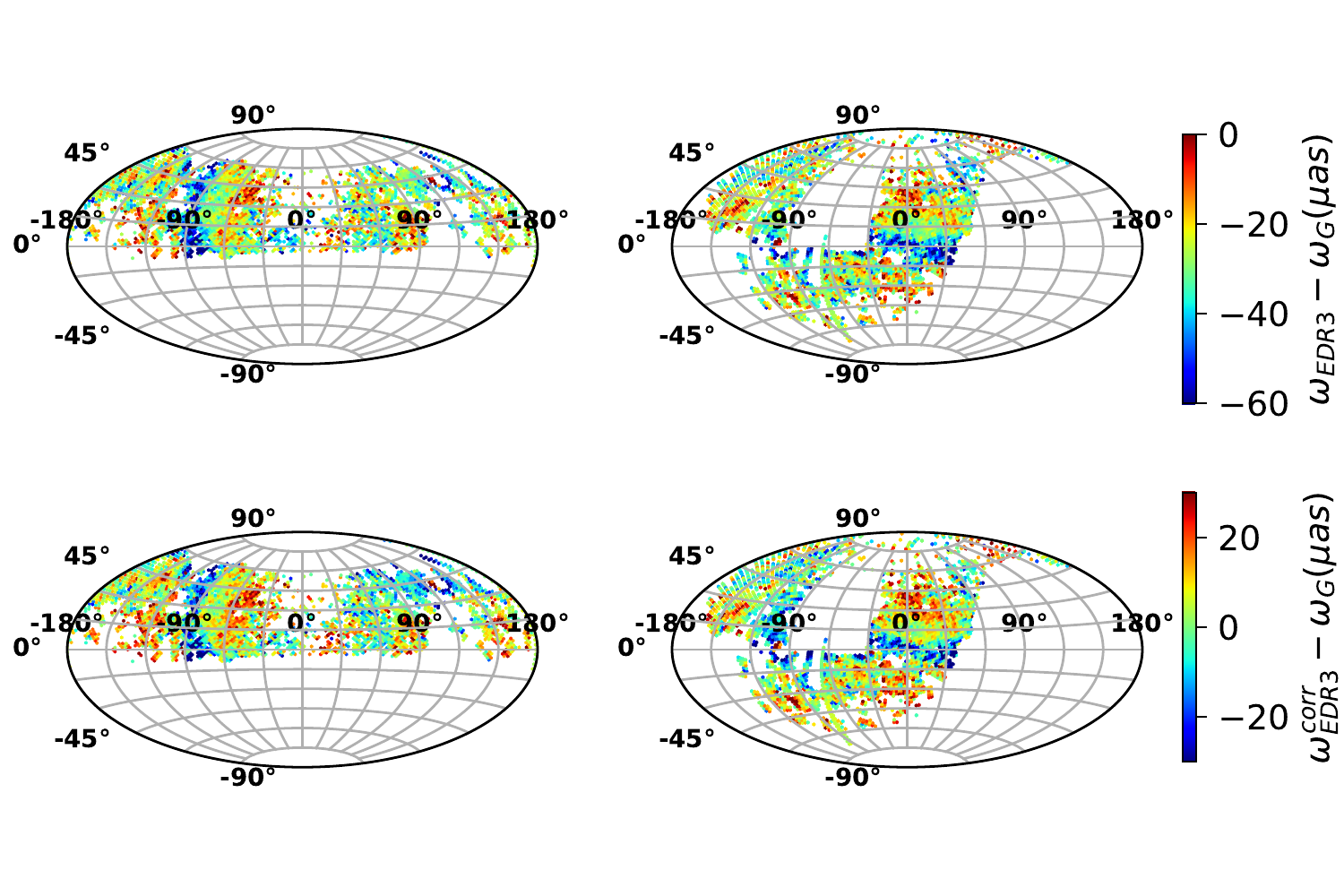}
\caption{Maps  of $\omega_{\rm EDR3}-\omega_{\rm G}$ (top panels) and $\omega_{\rm EDR3}^{\rm corr}-\omega_{\rm G}$ (bottom panels) in   equatorial coordinate (left panels) and Galactic coordinate (right panels) for the five-parameter solutions.Each pixel covers equal sky area of about $3.36 $ square degrees. Number of stars in each pixel is larger than 50. }
\label{zeropoint_lb}
\end{figure*}

 The  bias of Gaia EDR3 parallax is positional dependent \citep{Lindegren2021b, Huang2021, Ren2021}. Our LAMOST giant sample covers a large and continue volume of the Galaxy and has huge number of stars, thus it is a better sample to investigate the distribution of the parallax biases as a function of spatial positions. 
Fig.\,\ref{zeropoint_lb} shows the maps of the mean parallax differences  for the five-parameter solutions of $\omega_{\rm EDR3}-\omega_{\rm G}$ and $\omega_{\rm EDR3}^{\rm corr}-\omega_{\rm G}$ in equatorial coordinates and Galactic coordinates. The $\omega_{\rm EDR3}-\omega_{\rm G}$ are  smaller than 0 at almost all position bins, with a median value of $\sim -30 \,\mu \rm as$.  The $\omega_{\rm EDR3}^{\rm corr}-\omega_{\rm G}$ are  smaller than $+30 \,\mu \rm as$ and larger than $-30\, \mu \rm as$ at almost all position bins,  with a median value of $\sim 0 \,\mu \rm as$.  The results suggest that the official corrections for Gaia EDR3 parallaxes are effective in reducing global offset in the Gaia parallaxes.  
From this plot, we find  that the Gaia EDR3 parallax bias exhibit significant and clear trend with the positions. The parallax biases in the Galactic plane ($b \sim 0^{\circ}$) are largest, which is similar with the result of \cite{Lindegren2021b} and \cite{Ren2021}. 
Fig.\,\ref{zeropoint_lb}  suggests  that the parallax biases before and after correction show similar trends with the positions, which suggest that official correction for Gaia EDR3 parallaxes could not reduce the parallax bias variations with positions.   A spatially dependent  Gaia EDR3 parallax zero-point correction model is needed. 

We will provide a dataset, which contains the position-dependent Gaia EDR3 parallax zero-point corrections for five-parameter solution within the LAMOST footprint in the Galactic coordinate and equatorial coordinate.   The resolution of each pixel  is  about 3.36\,$\rm deg^{2}$ in HEALPix \citep[Hierarchical Equal Area isoLatitude Pixelisation;][]{Gorski2005} grid after dividing the whole sky into 12288  equal area pixels, which is same with that in Fig.\,\ref{zeropoint_lb}.  
For the stars located in a pixel we obtain the mean ($\mu$) and standard deviation ($\sigma$) of $\omega_{\rm EDR3}-\omega_{\rm G}$  and $\omega_{\rm EDR3}^{\rm corr}-\omega_{\rm G}$ by fitting a gaussian function. The dataset gives for each pixel $\mu$ as the Gaia EDR3 parallax zero-point correction in the pixel, $\frac{\sigma}{\sqrt{N}}$ as the corresponding error, and N as the number of stars in the pixel.
One can use this dataset to correct the spatially dependent Gaia EDR3 parallax zero-point offset. The dataset will  be released in the website of \url{http://www.lamost.org/dr8/v1.0/doc/vac}. Table.\,\ref{table1} shows the detailed description of the dataset. It is noted that we give both the position-dependent Gaia EDR3 parallax zero-point corrections in the Galactic and equatorial coordinates, one can choose one of them according to their coordinates.

\begin{deluxetable}{ccc}
\tablecaption{Descriptions for the dataset to correct spatially dependent Gaia EDR3 parallax zero-point offset for five-parameter solutions.
\label{table1}}
\tablehead{
\colhead{Field} & \colhead{Description} & \colhead{Unit}
}
\startdata
hpindex& the HEALPix index number (level  5)& -\\
ra&Right ascension at the center of each pixel &degree  \\
 dec&Declination at the center of each pixel& degree\\
 gl&Galactic longitude at the center of each pixel &degree \\
 gb& Galactic latitude at the center of each pixel&degree \\
 $\Delta \omega^{\rm corr}$&$\omega_{\rm EDR3}^{\rm corr}-\omega_{\rm G}$ values in each pixel& $\mu$as \\
 E$\_\Delta \omega^{\rm corr}$& the error of estimated $\Delta \omega^{\rm corr}$ in each pixel  &$\mu$as\\
 $\Delta \omega^{\rm noncorr}$&$\omega_{\rm EDR3}-\omega_{\rm G}$ values in each pixel & $\mu$as \\
 E$\_\Delta \omega^{\rm noncorr}$& the error of estimated $\Delta \omega^{\rm noncorr}$ in each pixel  &$\mu$as\\
  $N$&Number of stars in each pixel  &- \\
\enddata
\end{deluxetable}


\section{Summary}
In the current paper, we use $\sim$ 0.3 million  giant stars selected from  the LAMOST  to investigate the zero-point offset of the Gaia EDR3 parallax $\omega_{\rm EDR3}$ and the Gaia EDR3 parallax after correction $\omega_{\rm EDR3}^{\rm corr}$ using the model of  \cite{Lindegren2021b}.  The distance accuracy of these giant stars are about 8.5\%.  The global median zero-point offset of $\omega_{\rm EDR3}$  are  $-27.9 \,\mu \rm as$ and $-26.5 \,\mu \rm as$  for five-parameter solution and six-parameter solution, respectively. The bias of $\omega_{\rm EDR3}^{\rm corr}$ are $+$2.6\,$\mu \rm as$ and $+$2.9\,$\mu \rm as$ for five-parameter solution and six-parameter solution, respectively.

We also studied the main dependences of the Gaia EDR3 parallax bias  before and after correction.   We find that the parallax bias show  significant systematic trends with $G$ magnitude, spectral colors. The official parallax zero-point correction could reduce the global offfset in the Gaia EDR3 parallaxes and the significant systematic trends with $G$ magnitude. While the official parallax zero-point correction could not reduce the parallax bias variations with spectral colors. With the advantage of the huge number of stars in our giant stellar sample, we also investigate the parallax bias variations with positions.  Both the parallax bias before and after correction exhibit significant and clear trend with the positions, which suggest that  the official parallax zero-point correction could not reduce the parallax bias variations with positions. In the current paper, a spatially dependent  Gaia EDR3 parallax zero-point correction model for five-parameter solutions  in the LAMOST footprint is firstly provided.

\begin{acknowledgments}
This work was funded by the National Key R\&D Program of China
(No. 2019YFA0405500) and the National Natural Science Foundation of China (NSFC Grant No.11833006, U1531244 and 11973001, 12173007).  We used data from the European Space Agency mission Gaia (http://
www.cosmos.esa.int/gaia), processed by the Gaia Data Processing
and Analysis Consortium (DPAC; see http://www.cosmos.esa.
int/web/gaia/dpac/consortium). Guoshoujing Telescope (the Large Sky Area Multi-Object Fiber Spectroscopic Telescope LAMOST) is a National Major Scientific Project built by the Chinese Academy of Sciences. Funding for the project has been provided by the National Development and Reform Commission. LAMOST is operated and managed by the National Astronomical Observatories, Chinese Academy of Sciences.
\end{acknowledgments}

\bibliography{gaiaedr3}{}

\begin{thebibliography}{}
\expandafter\ifx\csname natexlab\endcsname\relax\def\natexlab#1{#1}\fi
\providecommand{\url}[1]{\href{#1}{#1}}
\providecommand{\dodoi}[1]{doi:~\href{http://doi.org/#1}{\nolinkurl{#1}}}
\providecommand{\doeprint}[1]{\href{http://ascl.net/#1}{\nolinkurl{http://ascl.net/#1}}}
\providecommand{\doarXiv}[1]{\href{https://arxiv.org/abs/#1}{\nolinkurl{https://arxiv.org/abs/#1}}}

\bibitem[{{Bailer-Jones} {et~al.}(2021){Bailer-Jones}, {Rybizki}, {Fouesneau},
  {Demleitner}, \& {Andrae}}]{Baileredr3}
{Bailer-Jones}, C.~A.~L., {Rybizki}, J., {Fouesneau}, M., {Demleitner}, M., \&
  {Andrae}, R. 2021, VizieR Online Data Catalog, I/352

\bibitem[{{Bhardwaj} {et~al.}(2021){Bhardwaj}, {Rejkuba}, {de Grijs}, {Yang},
  {Herczeg}, {Marconi}, {Singh}, {Kanbur}, \& {Ngeow}}]{Bhardwaj2021}
{Bhardwaj}, A., {Rejkuba}, M., {de Grijs}, R., {et~al.} 2021, \apj, 909, 200,
  \dodoi{10.3847/1538-4357/abdf48}

\bibitem[{{Bovy} {et~al.}(2014){Bovy}, {Nidever}, {Rix}, {Girardi}, {Zasowski},
  {Chojnowski}, {Holtzman}, {Epstein}, {Frinchaboy}, {Hayden}, {Rodrigues},
  {Majewski}, {Johnson}, {Pinsonneault}, {Stello}, {Allende Prieto}, {Andrews},
  {Basu}, {Beers}, {Bizyaev}, {Burton}, {Chaplin}, {Cunha}, {Elsworth},
  {Garc{\'\i}a}, {Garc{\'\i}a-Her{\'n}andez}, {Garc{\'\i}a P{\'e}rez},
  {Hearty}, {Hekker}, {Kallinger}, {Kinemuchi}, {Koesterke},
  {M{\'e}sz{\'a}ros}, {Mosser}, {O'Connell}, {Oravetz}, {Pan}, {Robin},
  {Schiavon}, {Schneider}, {Schultheis}, {Serenelli}, {Shetrone}, {Silva
  Aguirre}, {Simmons}, {Skrutskie}, {Smith}, {Stassun}, {Weinberg}, {Wilson},
  \& {Zamora}}]{Bovy2014}
{Bovy}, J., {Nidever}, D.~L., {Rix}, H.-W., {et~al.} 2014, \apj, 790, 127,
  \dodoi{10.1088/0004-637X/790/2/127}

\bibitem[{{Cannon}(1970)}]{Cannon1970}
{Cannon}, R.~D. 1970, \mnras, 150, 111, \dodoi{10.1093/mnras/150.1.111}

\bibitem[{{Chen} {et~al.}(2017){Chen}, {Casagrande}, {Zhao}, {Bovy}, {Silva
  Aguirre}, {Zhao}, \& {Jia}}]{Chen2017}
{Chen}, Y.~Q., {Casagrande}, L., {Zhao}, G., {et~al.} 2017, \apj, 840, 77,
  \dodoi{10.3847/1538-4357/aa6d0f}

\bibitem[{{Deng} {et~al.}(2012){Deng}, {Newberg}, {Liu}, {Carlin}, {Beers},
  {Chen}, {Chen}, {Christlieb}, {Grillmair}, {Guhathakurta}, {Han}, {Hou},
  {Lee}, {L{\'e}pine}, {Li}, {Liu}, {Pan}, {Sellwood}, {Wang}, {Wang}, {Yang},
  {Yanny}, {Zhang}, {Zhang}, {Zheng}, \& {Zhu}}]{deng-legue}
{Deng}, L.-C., {Newberg}, H.~J., {Liu}, C., {et~al.} 2012, Research in
  Astronomy and Astrophysics, 12, 735, \dodoi{10.1088/1674-4527/12/7/003}

\bibitem[{{El-Badry} {et~al.}(2021){El-Badry}, {Rix}, \&
  {Heintz}}]{ElBadry2021}
{El-Badry}, K., {Rix}, H.-W., \& {Heintz}, T.~M. 2021, \mnras, 506, 2269,
  \dodoi{10.1093/mnras/stab323}

\bibitem[{{Fabricius} {et~al.}(2021){Fabricius}, {Luri}, {Arenou}, {Babusiaux},
  {Helmi}, {Muraveva}, {Reyl{\'e}}, {Spoto}, {Vallenari}, {Antoja}, {Balbinot},
  {Barache}, {Bauchet}, {Bragaglia}, {Busonero}, {Cantat-Gaudin}, {Carrasco},
  {Diakit{\'e}}, {Fabrizio}, {Figueras}, {Garcia-Gutierrez}, {Garofalo},
  {Jordi}, {Kervella}, {Khanna}, {Leclerc}, {Licata}, {Lambert}, {Marrese},
  {Masip}, {Ramos}, {Robichon}, {Robin}, {Romero-G{\'o}mez}, {Rubele}, \&
  {Weiler}}]{Fabricius2021}
{Fabricius}, C., {Luri}, X., {Arenou}, F., {et~al.} 2021, \aap, 649, A5,
  \dodoi{10.1051/0004-6361/202039834}

\bibitem[{{Gaia Collaboration} {et~al.}(2016){Gaia Collaboration}, {Prusti},
  {de Bruijne}, {Brown}, {Vallenari}, {Babusiaux}, {Bailer-Jones}, {Bastian},
  {Biermann}, {Evans}, {Eyer}, {Jansen}, {Jordi}, {Klioner}, {Lammers},
  {Lindegren}, {Luri}, {Mignard}, {Milligan}, {Panem}, {Poinsignon},
  {Pourbaix}, {Randich}, {Sarri}, {Sartoretti}, {Siddiqui}, {Soubiran},
  {Valette}, {van Leeuwen}, {Walton}, {Aerts}, {Arenou}, {Cropper}, {Drimmel},
  {H{\o}g}, {Katz}, {Lattanzi}, {O'Mullane}, {Grebel}, {Holland}, {Huc},
  {Passot}, {Bramante}, {Cacciari}, {Casta{\~n}eda}, {Chaoul}, {Cheek}, {De
  Angeli}, {Fabricius}, {Guerra}, {Hern{\'a}ndez}, {Jean-Antoine-Piccolo},
  {Masana}, {Messineo}, {Mowlavi}, {Nienartowicz}, {Ord{\'o}{\~n}ez-Blanco},
  {Panuzzo}, {Portell}, {Richards}, {Riello}, {Seabroke}, {Tanga},
  {Th{\'e}venin}, {Torra}, {Els}, {Gracia-Abril}, {Comoretto},
  {Garcia-Reinaldos}, {Lock}, {Mercier}, {Altmann}, {Andrae}, {Astraatmadja},
  {Bellas-Velidis}, {Benson}, {Berthier}, {Blomme}, {Busso}, {Carry},
  {Cellino}, {Clementini}, {Cowell}, {Creevey}, {Cuypers}, {Davidson}, {De
  Ridder}, {de Torres}, {Delchambre}, {Dell'Oro}, {Ducourant}, {Fr{\'e}mat},
  {Garc{\'\i}a-Torres}, {Gosset}, {Halbwachs}, {Hambly}, {Harrison}, {Hauser},
  {Hestroffer}, {Hodgkin}, {Huckle}, {Hutton}, {Jasniewicz}, {Jordan},
  {Kontizas}, {Korn}, {Lanzafame}, {Manteiga}, {Moitinho}, {Muinonen},
  {Osinde}, {Pancino}, {Pauwels}, {Petit}, {Recio-Blanco}, {Robin}, {Sarro},
  {Siopis}, {Smith}, {Smith}, {Sozzetti}, {Thuillot}, {van Reeven}, {Viala},
  {Abbas}, {Abreu Aramburu}, {Accart}, {Aguado}, {Allan}, {Allasia},
  {Altavilla}, {{\'A}lvarez}, {Alves}, {Anderson}, {Andrei}, {Anglada Varela},
  {Antiche}, {Antoja}, {Ant{\'o}n}, {Arcay}, {Atzei}, {Ayache}, {Bach},
  {Baker}, {Balaguer-N{\'u}{\~n}ez}, {Barache}, {Barata}, {Barbier}, {Barblan},
  {Baroni}, {Barrado y Navascu{\'e}s}, {Barros}, {Barstow}, {Becciani},
  {Bellazzini}, {Bellei}, {Bello Garc{\'\i}a}, {Belokurov}, {Bendjoya},
  {Berihuete}, {Bianchi}, {Bienaym{\'e}}, {Billebaud}, {Blagorodnova},
  {Blanco-Cuaresma}, {Boch}, {Bombrun}, {Borrachero}, {Bouquillon}, {Bourda},
  {Bouy}, {Bragaglia}, {Breddels}, {Brouillet}, {Br{\"u}semeister},
  {Bucciarelli}, {Budnik}, {Burgess}, {Burgon}, {Burlacu}, {Busonero}, {Buzzi},
  {Caffau}, {Cambras}, {Campbell}, {Cancelliere}, {Cantat-Gaudin}, {Carlucci},
  {Carrasco}, {Castellani}, {Charlot}, {Charnas}, {Charvet}, {Chassat},
  {Chiavassa}, {Clotet}, {Cocozza}, {Collins}, {Collins}, {Costigan}, {Crifo},
  {Cross}, {Crosta}, {Crowley}, {Dafonte}, {Damerdji}, {Dapergolas}, {David},
  {David}, {De Cat}, {de Felice}, {de Laverny}, {De Luise}, {De March}, {de
  Martino}, {de Souza}, {Debosscher}, {del Pozo}, {Delbo}, {Delgado},
  {Delgado}, {di Marco}, {Di Matteo}, {Diakite}, {Distefano}, {Dolding}, {Dos
  Anjos}, {Drazinos}, {Dur{\'a}n}, {Dzigan}, {Ecale}, {Edvardsson}, {Enke},
  {Erdmann}, {Escolar}, {Espina}, {Evans}, {Eynard Bontemps}, {Fabre},
  {Fabrizio}, {Faigler}, {Falc{\~a}o}, {Farr{\`a}s Casas}, {Faye}, {Federici},
  {Fedorets}, {Fern{\'a}ndez-Hern{\'a}ndez}, {Fernique}, {Fienga}, {Figueras},
  {Filippi}, {Findeisen}, {Fonti}, {Fouesneau}, {Fraile}, {Fraser}, {Fuchs},
  {Furnell}, {Gai}, {Galleti}, {Galluccio}, {Garabato}, {Garc{\'\i}a-Sedano},
  {Gar{\'e}}, {Garofalo}, {Garralda}, {Gavras}, {Gerssen}, {Geyer}, {Gilmore},
  {Girona}, {Giuffrida}, {Gomes}, {Gonz{\'a}lez-Marcos},
  {Gonz{\'a}lez-N{\'u}{\~n}ez}, {Gonz{\'a}lez-Vidal}, {Granvik}, {Guerrier},
  {Guillout}, {Guiraud}, {G{\'u}rpide}, {Guti{\'e}rrez-S{\'a}nchez}, {Guy},
  {Haigron}, {Hatzidimitriou}, {Haywood}, {Heiter}, {Helmi}, {Hobbs},
  {Hofmann}, {Holl}, {Holland}, {Hunt}, {Hypki}, {Icardi}, {Irwin}, {Jevardat
  de Fombelle}, {Jofr{\'e}}, {Jonker}, {Jorissen}, {Julbe}, {Karampelas},
  {Kochoska}, {Kohley}, {Kolenberg}, {Kontizas}, {Koposov}, {Kordopatis},
  {Koubsky}, {Kowalczyk}, {Krone-Martins}, {Kudryashova}, {Kull}, {Bachchan},
  {Lacoste-Seris}, {Lanza}, {Lavigne}, {Le Poncin-Lafitte}, {Lebreton},
  {Lebzelter}, {Leccia}, {Leclerc}, {Lecoeur-Taibi}, {Lemaitre}, {Lenhardt},
  {Leroux}, {Liao}, {Licata}, {Lindstr{\o}m}, {Lister}, {Livanou}, {Lobel},
  {L{\"o}ffler}, {L{\'o}pez}, {Lopez-Lozano}, {Lorenz}, {Loureiro},
  {MacDonald}, {Magalh{\~a}es Fernandes}, {Managau}, {Mann}, {Mantelet},
  {Marchal}, {Marchant}, {Marconi}, {Marie}, {Marinoni}, {Marrese},
  {Marschalk{\'o}}, {Marshall}, {Mart{\'\i}n-Fleitas}, {Martino}, {Mary},
  {Matijevi{\v{c}}}, {Mazeh}, {McMillan}, {Messina}, {Mestre}, {Michalik},
  {Millar}, {Miranda}, {Molina}, {Molinaro}, {Molinaro}, {Moln{\'a}r},
  {Moniez}, {Montegriffo}, {Monteiro}, {Mor}, {Mora}, {Morbidelli}, {Morel},
  {Morgenthaler}, {Morley}, {Morris}, {Mulone}, {Muraveva}, {Musella},
  {Narbonne}, {Nelemans}, {Nicastro}, {Noval}, {Ord{\'e}novic},
  {Ordieres-Mer{\'e}}, {Osborne}, {Pagani}, {Pagano}, {Pailler}, {Palacin},
  {Palaversa}, {Parsons}, {Paulsen}, {Pecoraro}, {Pedrosa}, {Pentik{\"a}inen},
  {Pereira}, {Pichon}, {Piersimoni}, {Pineau}, {Plachy}, {Plum}, {Poujoulet},
  {Pr{\v{s}}a}, {Pulone}, {Ragaini}, {Rago}, {Rambaux}, {Ramos-Lerate},
  {Ranalli}, {Rauw}, {Read}, {Regibo}, {Renk}, {Reyl{\'e}}, {Ribeiro},
  {Rimoldini}, {Ripepi}, {Riva}, {Rixon}, {Roelens}, {Romero-G{\'o}mez},
  {Rowell}, {Royer}, {Rudolph}, {Ruiz-Dern}, {Sadowski}, {Sagrist{\`a}
  Sell{\'e}s}, {Sahlmann}, {Salgado}, {Salguero}, {Sarasso}, {Savietto},
  {Schnorhk}, {Schultheis}, {Sciacca}, {Segol}, {Segovia}, {Segransan},
  {Serpell}, {Shih}, {Smareglia}, {Smart}, {Smith}, {Solano}, {Solitro},
  {Sordo}, {Soria Nieto}, {Souchay}, {Spagna}, {Spoto}, {Stampa}, {Steele},
  {Steidelm{\"u}ller}, {Stephenson}, {Stoev}, {Suess}, {S{\"u}veges}, {Surdej},
  {Szabados}, {Szegedi-Elek}, {Tapiador}, {Taris}, {Tauran}, {Taylor},
  {Teixeira}, {Terrett}, {Tingley}, {Trager}, {Turon}, {Ulla}, {Utrilla},
  {Valentini}, {van Elteren}, {Van Hemelryck}, {van Leeuwen}, {Varadi},
  {Vecchiato}, {Veljanoski}, {Via}, {Vicente}, {Vogt}, {Voss}, {Votruba},
  {Voutsinas}, {Walmsley}, {Weiler}, {Weingrill}, {Werner}, {Wevers},
  {Whitehead}, {Wyrzykowski}, {Yoldas}, {{\v{Z}}erjal}, {Zucker}, {Zurbach},
  {Zwitter}, {Alecu}, {Allen}, {Allende Prieto}, {Amorim},
  {Anglada-Escud{\'e}}, {Arsenijevic}, {Azaz}, {Balm}, {Beck}, {Bernstein},
  {Bigot}, {Bijaoui}, {Blasco}, {Bonfigli}, {Bono}, {Boudreault}, {Bressan},
  {Brown}, {Brunet}, {Bunclark}, {Buonanno}, {Butkevich}, {Carret}, {Carrion},
  {Chemin}, {Ch{\'e}reau}, {Corcione}, {Darmigny}, {de Boer}, {de Teodoro}, {de
  Zeeuw}, {Delle Luche}, {Domingues}, {Dubath}, {Fodor}, {Fr{\'e}zouls},
  {Fries}, {Fustes}, {Fyfe}, {Gallardo}, {Gallegos}, {Gardiol}, {Gebran},
  {Gomboc}, {G{\'o}mez}, {Grux}, {Gueguen}, {Heyrovsky}, {Hoar}, {Iannicola},
  {Isasi Parache}, {Janotto}, {Joliet}, {Jonckheere}, {Keil}, {Kim},
  {Klagyivik}, {Klar}, {Knude}, {Kochukhov}, {Kolka}, {Kos}, {Kutka}, {Lainey},
  {LeBouquin}, {Liu}, {Loreggia}, {Makarov}, {Marseille}, {Martayan},
  {Martinez-Rubi}, {Massart}, {Meynadier}, {Mignot}, {Munari}, {Nguyen},
  {Nordlander}, {Ocvirk}, {O'Flaherty}, {Olias Sanz}, {Ortiz}, {Osorio},
  {Oszkiewicz}, {Ouzounis}, {Palmer}, {Park}, {Pasquato}, {Peltzer}, {Peralta},
  {P{\'e}turaud}, {Pieniluoma}, {Pigozzi}, {Poels}, {Prat}, {Prod'homme},
  {Raison}, {Rebordao}, {Risquez}, {Rocca-Volmerange}, {Rosen}, {Ruiz-Fuertes},
  {Russo}, {Sembay}, {Serraller Vizcaino}, {Short}, {Siebert}, {Silva},
  {Sinachopoulos}, {Slezak}, {Soffel}, {Sosnowska}, {Strai{\v{z}}ys}, {ter
  Linden}, {Terrell}, {Theil}, {Tiede}, {Troisi}, {Tsalmantza}, {Tur},
  {Vaccari}, {Vachier}, {Valles}, {Van Hamme}, {Veltz}, {Virtanen}, {Wallut},
  {Wichmann}, {Wilkinson}, {Ziaeepour}, \& {Zschocke}}]{Gaia_Collaboration2016}
{Gaia Collaboration}, {Prusti}, T., {de Bruijne}, J.~H.~J., {et~al.} 2016,
  \aap, 595, A1, \dodoi{10.1051/0004-6361/201629272}

\bibitem[{{Gaia Collaboration} {et~al.}(2021){Gaia Collaboration}, {Brown},
  {Vallenari}, {Prusti}, {de Bruijne}, {Babusiaux}, {Biermann}, {Creevey},
  {Evans}, {Eyer}, {Hutton}, {Jansen}, {Jordi}, {Klioner}, {Lammers},
  {Lindegren}, {Luri}, {Mignard}, {Panem}, {Pourbaix}, {Randich}, {Sartoretti},
  {Soubiran}, {Walton}, {Arenou}, {Bailer-Jones}, {Bastian}, {Cropper},
  {Drimmel}, {Katz}, {Lattanzi}, {van Leeuwen}, {Bakker}, {Cacciari},
  {Casta{\~n}eda}, {De Angeli}, {Ducourant}, {Fabricius}, {Fouesneau},
  {Fr{\'e}mat}, {Guerra}, {Guerrier}, {Guiraud}, {Jean-Antoine Piccolo},
  {Masana}, {Messineo}, {Mowlavi}, {Nicolas}, {Nienartowicz}, {Pailler},
  {Panuzzo}, {Riclet}, {Roux}, {Seabroke}, {Sordo}, {Tanga}, {Th{\'e}venin},
  {Gracia-Abril}, {Portell}, {Teyssier}, {Altmann}, {Andrae}, {Bellas-Velidis},
  {Benson}, {Berthier}, {Blomme}, {Brugaletta}, {Burgess}, {Busso}, {Carry},
  {Cellino}, {Cheek}, {Clementini}, {Damerdji}, {Davidson}, {Delchambre},
  {Dell'Oro}, {Fern{\'a}ndez-Hern{\'a}ndez}, {Galluccio}, {Garc{\'\i}a-Lario},
  {Garcia-Reinaldos}, {Gonz{\'a}lez-N{\'u}{\~n}ez}, {Gosset}, {Haigron},
  {Halbwachs}, {Hambly}, {Harrison}, {Hatzidimitriou}, {Heiter},
  {Hern{\'a}ndez}, {Hestroffer}, {Hodgkin}, {Holl}, {Jan{\ss}en}, {Jevardat de
  Fombelle}, {Jordan}, {Krone-Martins}, {Lanzafame}, {L{\"o}ffler}, {Lorca},
  {Manteiga}, {Marchal}, {Marrese}, {Moitinho}, {Mora}, {Muinonen}, {Osborne},
  {Pancino}, {Pauwels}, {Petit}, {Recio-Blanco}, {Richards}, {Riello},
  {Rimoldini}, {Robin}, {Roegiers}, {Rybizki}, {Sarro}, {Siopis}, {Smith},
  {Sozzetti}, {Ulla}, {Utrilla}, {van Leeuwen}, {van Reeven}, {Abbas}, {Abreu
  Aramburu}, {Accart}, {Aerts}, {Aguado}, {Ajaj}, {Altavilla}, {{\'A}lvarez},
  {{\'A}lvarez Cid-Fuentes}, {Alves}, {Anderson}, {Anglada Varela}, {Antoja},
  {Audard}, {Baines}, {Baker}, {Balaguer-N{\'u}{\~n}ez}, {Balbinot}, {Balog},
  {Barache}, {Barbato}, {Barros}, {Barstow}, {Bartolom{\'e}}, {Bassilana},
  {Bauchet}, {Baudesson-Stella}, {Becciani}, {Bellazzini}, {Bernet}, {Bertone},
  {Bianchi}, {Blanco-Cuaresma}, {Boch}, {Bombrun}, {Bossini}, {Bouquillon},
  {Bragaglia}, {Bramante}, {Breedt}, {Bressan}, {Brouillet}, {Bucciarelli},
  {Burlacu}, {Busonero}, {Butkevich}, {Buzzi}, {Caffau}, {Cancelliere},
  {C{\'a}novas}, {Cantat-Gaudin}, {Carballo}, {Carlucci}, {Carnerero},
  {Carrasco}, {Casamiquela}, {Castellani}, {Castro-Ginard}, {Castro Sampol},
  {Chaoul}, {Charlot}, {Chemin}, {Chiavassa}, {Cioni}, {Comoretto}, {Cooper},
  {Cornez}, {Cowell}, {Crifo}, {Crosta}, {Crowley}, {Dafonte}, {Dapergolas},
  {David}, {David}, {de Laverny}, {De Luise}, {De March}, {De Ridder}, {de
  Souza}, {de Teodoro}, {de Torres}, {del Peloso}, {del Pozo}, {Delbo},
  {Delgado}, {Delgado}, {Delisle}, {Di Matteo}, {Diakite}, {Diener},
  {Distefano}, {Dolding}, {Eappachen}, {Edvardsson}, {Enke}, {Esquej}, {Fabre},
  {Fabrizio}, {Faigler}, {Fedorets}, {Fernique}, {Fienga}, {Figueras},
  {Fouron}, {Fragkoudi}, {Fraile}, {Franke}, {Gai}, {Garabato},
  {Garcia-Gutierrez}, {Garc{\'\i}a-Torres}, {Garofalo}, {Gavras}, {Gerlach},
  {Geyer}, {Giacobbe}, {Gilmore}, {Girona}, {Giuffrida}, {Gomel}, {Gomez},
  {Gonzalez-Santamaria}, {Gonz{\'a}lez-Vidal}, {Granvik},
  {Guti{\'e}rrez-S{\'a}nchez}, {Guy}, {Hauser}, {Haywood}, {Helmi}, {Hidalgo},
  {Hilger}, {H{\l}adczuk}, {Hobbs}, {Holland}, {Huckle}, {Jasniewicz},
  {Jonker}, {Juaristi Campillo}, {Julbe}, {Karbevska}, {Kervella}, {Khanna},
  {Kochoska}, {Kontizas}, {Kordopatis}, {Korn}, {Kostrzewa-Rutkowska},
  {Kruszy{\'n}ska}, {Lambert}, {Lanza}, {Lasne}, {Le Campion}, {Le Fustec},
  {Lebreton}, {Lebzelter}, {Leccia}, {Leclerc}, {Lecoeur-Taibi}, {Liao},
  {Licata}, {Lindstr{\o}m}, {Lister}, {Livanou}, {Lobel}, {Madrero Pardo},
  {Managau}, {Mann}, {Marchant}, {Marconi}, {Marcos Santos}, {Marinoni},
  {Marocco}, {Marshall}, {Martin Polo}, {Mart{\'\i}n-Fleitas}, {Masip},
  {Massari}, {Mastrobuono-Battisti}, {Mazeh}, {McMillan}, {Messina},
  {Michalik}, {Millar}, {Mints}, {Molina}, {Molinaro}, {Moln{\'a}r},
  {Montegriffo}, {Mor}, {Morbidelli}, {Morel}, {Morris}, {Mulone}, {Munoz},
  {Muraveva}, {Murphy}, {Musella}, {Noval}, {Ord{\'e}novic}, {Orr{\`u}},
  {Osinde}, {Pagani}, {Pagano}, {Palaversa}, {Palicio}, {Panahi}, {Pawlak},
  {Pe{\~n}alosa Esteller}, {Penttil{\"a}}, {Piersimoni}, {Pineau}, {Plachy},
  {Plum}, {Poggio}, {Poretti}, {Poujoulet}, {Pr{\v{s}}a}, {Pulone}, {Racero},
  {Ragaini}, {Rainer}, {Raiteri}, {Rambaux}, {Ramos}, {Ramos-Lerate}, {Re
  Fiorentin}, {Regibo}, {Reyl{\'e}}, {Ripepi}, {Riva}, {Rixon}, {Robichon},
  {Robin}, {Roelens}, {Rohrbasser}, {Romero-G{\'o}mez}, {Rowell}, {Royer},
  {Rybicki}, {Sadowski}, {Sagrist{\`a} Sell{\'e}s}, {Sahlmann}, {Salgado},
  {Salguero}, {Samaras}, {Sanchez Gimenez}, {Sanna}, {Santove{\~n}a},
  {Sarasso}, {Schultheis}, {Sciacca}, {Segol}, {Segovia}, {S{\'e}gransan},
  {Semeux}, {Shahaf}, {Siddiqui}, {Siebert}, {Siltala}, {Slezak}, {Smart},
  {Solano}, {Solitro}, {Souami}, {Souchay}, {Spagna}, {Spoto}, {Steele},
  {Steidelm{\"u}ller}, {Stephenson}, {S{\"u}veges}, {Szabados}, {Szegedi-Elek},
  {Taris}, {Tauran}, {Taylor}, {Teixeira}, {Thuillot}, {Tonello}, {Torra},
  {Torra}, {Turon}, {Unger}, {Vaillant}, {van Dillen}, {Vanel}, {Vecchiato},
  {Viala}, {Vicente}, {Voutsinas}, {Weiler}, {Wevers}, {Wyrzykowski}, {Yoldas},
  {Yvard}, {Zhao}, {Zorec}, {Zucker}, {Zurbach}, \&
  {Zwitter}}]{Gaia_Collaboration2021}
{Gaia Collaboration}, {Brown}, A.~G.~A., {Vallenari}, A., {et~al.} 2021, \aap,
  649, A1, \dodoi{10.1051/0004-6361/202039657}

\bibitem[{{G{\'o}rski} {et~al.}(2005){G{\'o}rski}, {Hivon}, {Banday},
  {Wandelt}, {Hansen}, {Reinecke}, \& {Bartelmann}}]{Gorski2005}
{G{\'o}rski}, K.~M., {Hivon}, E., {Banday}, A.~J., {et~al.} 2005, \apj, 622,
  759, \dodoi{10.1086/427976}

\bibitem[{{Groenewegen}(2021)}]{Groenewegen2021}
{Groenewegen}, M. 2021, arXiv e-prints, arXiv:2106.08128.
\newblock \doarXiv{2106.08128}

\bibitem[{{Huang} {et~al.}(2021){Huang}, {Yuan}, {Beers}, \&
  {Zhang}}]{Huang2021}
{Huang}, Y., {Yuan}, H., {Beers}, T.~C., \& {Zhang}, H. 2021, \apjl, 910, L5,
  \dodoi{10.3847/2041-8213/abe69a}

\bibitem[{{Huang} {et~al.}(2015){Huang}, {Liu}, {Zhang}, {Yuan}, {Xiang},
  {Chen}, {Ren}, {Sun}, {Wang}, {Zhang}, {Hou}, {Wang}, \& {Yang}}]{Huang2015}
{Huang}, Y., {Liu}, X.-W., {Zhang}, H.-W., {et~al.} 2015, Research in Astronomy
  and Astrophysics, 15, 1240, \dodoi{10.1088/1674-4527/15/8/010}

\bibitem[{{Huang} {et~al.}(2020){Huang}, {Sch{\"o}nrich}, {Zhang}, {Wu},
  {Chen}, {Wang}, {Xiang}, {Wang}, {Yuan}, {Li}, {Sun}, {Li}, \&
  {Liu}}]{Huang2020}
{Huang}, Y., {Sch{\"o}nrich}, R., {Zhang}, H., {et~al.} 2020, \apjs, 249, 29,
  \dodoi{10.3847/1538-4365/ab994f}

\bibitem[{{Liao} {et~al.}(2021){Liao}, {Wu}, {Qi}, {Tang}, {Luo}, \&
  {Cao}}]{Liao2021}
{Liao}, S., {Wu}, Q., {Qi}, Z., {et~al.} 2021, \pasp, 133, 094501,
  \dodoi{10.1088/1538-3873/ac1eeb}

\bibitem[{{Lindegren} {et~al.}(2021{\natexlab{a}}){Lindegren}, {Klioner},
  {Hern{\'a}ndez}, {Bombrun}, {Ramos-Lerate}, {Steidelm{\"u}ller}, {Bastian},
  {Biermann}, {de Torres}, {Gerlach}, {Geyer}, {Hilger}, {Hobbs}, {Lammers},
  {McMillan}, {Stephenson}, {Casta{\~n}eda}, {Davidson}, {Fabricius},
  {Gracia-Abril}, {Portell}, {Rowell}, {Teyssier}, {Torra}, {Bartolom{\'e}},
  {Clotet}, {Garralda}, {Gonz{\'a}lez-Vidal}, {Torra}, {Abbas}, {Altmann},
  {Anglada Varela}, {Balaguer-N{\'u}{\~n}ez}, {Balog}, {Barache}, {Becciani},
  {Bernet}, {Bertone}, {Bianchi}, {Bouquillon}, {Brown}, {Bucciarelli},
  {Busonero}, {Butkevich}, {Buzzi}, {Cancelliere}, {Carlucci}, {Charlot},
  {Cioni}, {Crosta}, {Crowley}, {del Peloso}, {del Pozo}, {Drimmel}, {Esquej},
  {Fienga}, {Fraile}, {Gai}, {Garcia-Reinaldos}, {Guerra}, {Hambly}, {Hauser},
  {Jan{\ss}en}, {Jordan}, {Kostrzewa-Rutkowska}, {Lattanzi}, {Liao}, {Licata},
  {Lister}, {L{\"o}ffler}, {Marchant}, {Masip}, {Mignard}, {Mints}, {Molina},
  {Mora}, {Morbidelli}, {Murphy}, {Pagani}, {Panuzzo}, {Pe{\~n}alosa Esteller},
  {Poggio}, {Re Fiorentin}, {Riva}, {Sagrist{\`a} Sell{\'e}s}, {Sanchez
  Gimenez}, {Sarasso}, {Sciacca}, {Siddiqui}, {Smart}, {Souami}, {Spagna},
  {Steele}, {Taris}, {Utrilla}, {van Reeven}, \& {Vecchiato}}]{Lindegren2021}
{Lindegren}, L., {Klioner}, S.~A., {Hern{\'a}ndez}, J., {et~al.}
  2021{\natexlab{a}}, \aap, 649, A2, \dodoi{10.1051/0004-6361/202039709}

\bibitem[{{Lindegren} {et~al.}(2021{\natexlab{b}}){Lindegren}, {Bastian},
  {Biermann}, {Bombrun}, {de Torres}, {Gerlach}, {Geyer}, {Hern{\'a}ndez},
  {Hilger}, {Hobbs}, {Klioner}, {Lammers}, {McMillan}, {Ramos-Lerate},
  {Steidelm{\"u}ller}, {Stephenson}, \& {van Leeuwen}}]{Lindegren2021b}
{Lindegren}, L., {Bastian}, U., {Biermann}, M., {et~al.} 2021{\natexlab{b}},
  \aap, 649, A4, \dodoi{10.1051/0004-6361/202039653}

\bibitem[{{Liu} {et~al.}(2014){Liu}, {Yuan}, {Huo}, {Deng}, {Hou}, {Zhao},
  {Zhao}, {Shi}, {Luo}, {Xiang}, {Zhang}, {Huang}, \& {Zhang}}]{liu-lss-gac}
{Liu}, X.-W., {Yuan}, H.-B., {Huo}, Z.-Y., {et~al.} 2014, in IAU Symposium,
  Vol. 298, IAU Symposium, ed. S.~{Feltzing}, G.~{Zhao}, N.~A. {Walton}, \&
  P.~{Whitelock}, 310--321, \dodoi{10.1017/S1743921313006510}

\bibitem[{{Paczy{\'n}ski} \& {Stanek}(1998)}]{Paczynski1998}
{Paczy{\'n}ski}, B., \& {Stanek}, K.~Z. 1998, \apjl, 494, L219,
  \dodoi{10.1086/311181}

\bibitem[{{Ren} {et~al.}(2021){Ren}, {Chen}, {Zhang}, {de Grijs}, {Deng}, \&
  {Huang}}]{Ren2021}
{Ren}, F., {Chen}, X., {Zhang}, H., {et~al.} 2021, \apjl, 911, L20,
  \dodoi{10.3847/2041-8213/abf359}

\bibitem[{{Skrutskie} {et~al.}(2006){Skrutskie}, {Cutri}, {Stiening},
  {Weinberg}, {Schneider}, {Carpenter}, {Beichman}, {Capps}, {Chester},
  {Elias}, {Huchra}, {Liebert}, {Lonsdale}, {Monet}, {Price}, {Seitzer},
  {Jarrett}, {Kirkpatrick}, {Gizis}, {Howard}, {Evans}, {Fowler}, {Fullmer},
  {Hurt}, {Light}, {Kopan}, {Marsh}, {McCallon}, {Tam}, {Van Dyk}, \&
  {Wheelock}}]{Skrutskie2006}
{Skrutskie}, M.~F., {Cutri}, R.~M., {Stiening}, R., {et~al.} 2006, \aj, 131,
  1163, \dodoi{10.1086/498708}

\bibitem[{{Stassun} \& {Torres}(2021)}]{Stassun2021}
{Stassun}, K.~G., \& {Torres}, G. 2021, \apjl, 907, L33,
  \dodoi{10.3847/2041-8213/abdaad}

\bibitem[{{Wan} {et~al.}(2015){Wan}, {Liu}, {Deng}, {Cui}, {Zhang}, {Hou},
  {Yang}, \& {Wu}}]{Wan2015}
{Wan}, J.-C., {Liu}, C., {Deng}, L.-C., {et~al.} 2015, Research in Astronomy
  and Astrophysics, 15, 1166, \dodoi{10.1088/1674-4527/15/8/006}

\bibitem[{{Yuan} {et~al.}(2013){Yuan}, {Liu}, \& {Xiang}}]{Yuan2013}
{Yuan}, H.~B., {Liu}, X.~W., \& {Xiang}, M.~S. 2013, \mnras, 430, 2188,
  \dodoi{10.1093/mnras/stt039}

\bibitem[{{Zhao} {et~al.}(2012){Zhao}, {Zhao}, {Chu}, {Jing}, \&
  {Deng}}]{Zhao2012}
{Zhao}, G., {Zhao}, Y.-H., {Chu}, Y.-Q., {Jing}, Y.-P., \& {Deng}, L.-C. 2012,
  Research in Astronomy and Astrophysics, 12, 723,
  \dodoi{10.1088/1674-4527/12/7/002}

\bibitem[{{Zinn}(2021)}]{Zinn2021}
{Zinn}, J.~C. 2021, \aj, 161, 214, \dodoi{10.3847/1538-3881/abe936}

\end{thebibliography}
\bibliographystyle{aasjournal}

\end{document}